\documentclass[aps,nofootinbib]{revtex4}
\usepackage{amsfonts,color,bbm,amsmath,amssymb}
\usepackage{graphicx,calc,epsfig}

\definecolor{blue}{rgb}{0,0,1}
\definecolor{green}{rgb}{0,1,0}
\definecolor{red}{rgb}{1,0,0}
\definecolor{van}{rgb}{1,0,1}
\definecolor{al}{rgb}{1,1,0}

\newcounter{mnotecount}[section]

\newcommand{\MM}{\mathbb{M}}

\newcommand{\be}{\nopagebreak[3]\begin{equation}}
\newcommand{\ee}{\end{equation}}
\newcommand{\ba}{\nopagebreak[3]\begin{eqnarray}}
\newcommand{\ea}{\end{eqnarray}}
\DeclareFontFamily{U}{rsfs}{}         
\DeclareFontShape{U}{rsfs}{m}{n}{<5> rsfs5 <6><7> rsfs7          %
  <8><9><10><10.95><12><14.4><17.28><20.74><24.88> rsfs10}{}     %
\DeclareMathAlphabet{\mathfs}{U}{rsfs}{m}{n}                     %
\newcommand{\mfs}[1]{\mathfs {#1}}                               %
\newcommand{\va}{\scriptscriptstyle}

\newcommand{\sL}{{\mfs L}}

\newcommand{\sM}{{\mfs M}}

\newcommand{\Pp}{{}^{ \va +}\!\!{\mfs P}}
\newcommand{\Pm}{{}^{ \va -}\!\!{\mfs P}}
\newcommand{\Ppm}{{}^{ \va \pm}\!\!{\mfs P}}
\newcommand{\op}{{^{\va +}\!\omega}}

\newcommand{\om}{^{\va -}\!\omega}

\usepackage{psfrag}
\usepackage{vmargin}
\usepackage{amsfonts}
\usepackage{amsmath}
\usepackage{graphics,graphicx,color,calc,epsfig}
\usepackage{euscript}
\newcommand{\beq}{\begin{equation}}
\newcommand{\eeq}{\end{equation}}
\newcommand{\beqa}{\begin{eqnarray}}
\newcommand{\eeqa}{\end{eqnarray}}

\newcommand{\R}{\mathbb{R}}

\setpapersize{USletter}
\setmarginsrb{39mm}{12mm}{28mm}{25mm}{12pt}{11mm}{0pt}{13mm}

\begin{document}

\title{4d Lorentzian Holst action with topological terms}

\author{Danilo Jimenez Rezende}
\affiliation{Centre de Physique Th\'eorique\footnote{Unit\'e Mixte
de Recherche (UMR 6207) du CNRS et des Universit\'es Aix-Marseille
I, Aix-Marseille II, et du Sud Toulon-Var; laboratoire afili\'e
\`a la FRUMAM (FR 2291)}, Campus de Luminy, 13288 Marseille,
France.}

\author{Alejandro Perez}
\affiliation{Centre de Physique Th\'eorique\footnote{Unit\'e Mixte
de Recherche (UMR 6207) du CNRS et des Universit\'es Aix-Marseille
I, Aix-Marseille II, et du Sud Toulon-Var; laboratoire afili\'e
\`a la FRUMAM (FR 2291)}, Campus de Luminy, 13288 Marseille,
France.}

\date{\today \vbox{\vskip 2em}}

\begin{abstract}

We study the Hamiltonian formulation of the general first order
action of general relativity compatible with local Lorentz invariance 
and background independence. The most general
simplectic structure (compatible with diffeomorphism invariance and
local Lorentz transformations) is obtained by adding to the Holst
action the Pontriagin, Euler and Nieh-Yan invariants with
independent coupling constants. We perform a detailed canonical
analysis of this general formulation (in the time gauge) exploring
the structure of the phase space in terms of connection variables.
We explain the relationship of these topological terms, and the
effect of large $SU(2)$ gauge transformations in quantum theories of
gravity defined in terms of the Ashtekar-Barbero connection.

\end{abstract}

\maketitle

\section{Introduction}

The possibility of describing the phase space of gravity as a
background independent $SU(2)$ connection gauge theory is a remarkable
property of the first order formulation of general relativity in four
dimensions. This is the basis of  the 
canonical quantization program of gravity known as loop quantum gravity \cite{lqg}. 
After the discovery of the self dual connection
formulation of canonical general relativity by Ashtekar \cite{ash1},
it was soon realized by Barbero \cite{barb} that a formulation in
terms of a real $SU(2)$ connection was indeed possible.  The only
price to be paid is the appearance of a new free parameter $\gamma\in
\R-\{0\}$ (the so-called Immirzi parameter \cite{Imm}) into the
definition of the canonical variables. A first step in clarifying the
origin of the Immirzi parameter was to show\cite{holst} that the
Ashtekar-Barbero variables can be obtained directly from the
Hamiltonian formulation of general relativity defined by the first
order action \be S[e,\omega] = \int \star \left( e^I \wedge e^J
\right) \wedge F_{J I}(\omega) + \frac{1}{\gamma} e^I \wedge e^J
\wedge F_{I J}(\omega)\label{one}\ee where $e$ is a vierbein and
$\omega$ is a Lorentz connection. The first term is the standard
Palatini action of general relativity, while second term can be shown
not to affect the classical equations of motion. The reason for this
is that $\delta_{\omega} S=0$ is independent of $\gamma$, and implies
the connection to be the uniquely defined torsion free connection
compatible with $e$: $\omega=\omega(e)$. The second term contribution to the equation 
$\delta_e S=0$ vanishes identically when evaluated on $\omega(e)$ due to the
Riemann tensor identity $R_{[abc]d}=0$.

The canonical formulation of the Holst action leads in this way to a
one parameter family of $SU(2)$ connection formulations of the phase
space of general relativity: all of them related by canonical
transformations.  However, in the quantum theory the canonical
transformations relating different connection formulations appear
not to be unitarily implemented. For instance the spectra of 
geometric operators are modulated by $\gamma$. Formally speaking, the off shell
contributions of the second term in the action (\ref{one}) have a
non trivial effect on amplitudes in the path integral formulation of
quantum gravity.

There is at least another real parameter---describing the family of
possible $SU(2)$ connection formulations of gravity---with very
similar qualitative effects: the so called $\theta$ parameter.  Again,
this parameter labels classically equivalent formulations that become
physically different upon quantization. The reason for this is
geometrically more transparent than the case of the Immirzi parameter
as the effects of the $\theta$ parameter in the quantum theory are
associated to the transformation properties of physical states under
large $SU(2)$ gauge transformations \cite{ashty, largegauge}.  Physically, the
phenomenon is in strict analogy with the $\theta$ parameter effects in
QCD. 

All this motivates the following questions (we shall explore in this
work): Are there yet more general $SU(2)$ connection formulations of
gravity? i.e., are there new parameters in addition to $\gamma$
and $\theta$? and if so, how naturally they arise from the Lagrangian
framework, and , what are their possible physical effects upon
quantization? More particularly, does the $\theta$ parameter in the
connection formulation of gravity have a natural description at the
Lagrangian level? We will shed some light on these questions by
studying the canonical formulation of a {\em general family} of
actions for general relativity (in a sense described below).

Holst's action allows to understand the presence of a non-vanishing and
finite Immirzi parameter from a more clear standpoint. In fact, not
having the second term in the first order formulation of general
relativity (i.e. choosing $\gamma=\infty$ or in other words the
Palatini formulation) would be un-natural from the Wilsonian
perspective that calls for including in the action principle all terms
compatible with the symmetry and field content of the theory.  From
this perspective the Immirzi parameter $\gamma\in \R-\{0\}$ is not an
input but a consequence of local Lorentz plus diffeomorphism
invariance together with the choice of $e$ and $\omega$ as fundamental
fields. If we pursue this logic further then the most general action
principle---compatible with diffeomorphism invariance and Lorentz
invariance---describing pure-gravity in the first order formalism is
\begin{widetext}
\begin{eqnarray}
  S[e,\omega] & = & \int \overbrace{ \alpha_1 \star \left( e^I \wedge
  e^J \right) \wedge F_{J I}(\omega) + \alpha_2 e^I \wedge e^J \wedge
  F_{I J}(\omega)}^{Holst} \nonumber + \overbrace{\alpha_3 F^{I
  J}(\omega) \wedge F_{J I}(\omega)}^{Pontrjagin}\\ \nonumber & +&
  \underbrace{\alpha_4 F^{I J}(\omega) \wedge \star F_{J
  I}(\omega)}_{Euler} + \alpha_5\ \underbrace{d_{\omega}e^I \wedge d_{\omega} e_I -e^I \wedge e^J \wedge
  F_{I J}(\omega)}_{Nieh-Yan}\\ &+&\alpha_6\
\underbrace{  e^I \wedge e^J\wedge e^K \wedge e^L\epsilon_{IJKL}}_{Cosmological\  constant} ,
  \label{action}
\end{eqnarray}
\end{widetext}
where $\alpha_1$, $\alpha_2$ and $\alpha_5$ have $M^2$ dimension,
$\alpha_3$ and $\alpha_4$ are real dimensionless parameters, and
$\alpha_6$ is proportional to the cosmological constant. It is a
remarkable feature that only finitely many terms are allowed by the
symmetry once first order variables are chosen. This is in
clear contrast with the formulation of pure-gravity in terms of
metric variables where the most general action has infinitely many
(higher-curvature) contributions\footnote{Incidentally, this would
imply the renormalizability of quantum gravity if in the
construction of the quantum theory one could find a regularization
prescription compatible with the symmetries of (\ref{action}).}. We
should point out that special case of the previous general action
have been studied in the literature by Montesinos \cite{merced} and
more recently by Date et al. in \cite{Date:2008rb} and by Mercuri in \cite{Mercuri:2006um} (the last two references consider coupling with fermion). We will discuss
in detail these special cases a the end of this paper.

Notice that the terms proportional to $\alpha_3$, $\alpha_4$, and
$\alpha_5$ are the Pontrjagin, the Euler, and the Nieh-Yan classes
respectively. As the term proportional to $\alpha_2$, these
topological invariants have no effect on the equations of motion of
gravity as they can be written as the exterior derivative of
suitable $3$-forms (see equation (\ref{LHPE}) below). However, these
boundary terms affect the canonical structure of the theory:
they act as generating functionals of canonical transformations. As mentioned
above this might have physical relevance when these canonical
transformations cannot be unitarily implemented in the quantum
theory. Heuristically, the off-shell contributions of the
topological terms in (\ref{action}) to transition amplitudes (in the
language of the functional integral) might have non trivial effect
in quantum gravity.

The paper is organized as follows: In the following section we
perform the canonical analysis of the action (\ref{action}). In
addition to providing a complete analysis of the effect of the
addition of topological invariants to the Holst action (completing
existing existing results in the literature \cite{merced,
Date:2008rb}), this section provides a detailed presentation of
Holst's results \cite{holst} in a way that is alternative to the formulation of Barros e
Sa \cite{sa}. As there are second class constraints
that, for general values of the couplings, cannot be explicitly
solved in Section \ref{dirac} we compute the Dirac brackets in all
generality. In Section \ref{cano} we specialize to the family of
couplings for which second class constraints can be solved, and we
show that the term leading to the $\theta$ parameter in gravity
cannot be obtained for real couplings. We discuss the way in which
the $\theta$ term can be introduced at the Lagrangian level in
Section \ref{thetas}. We conclude with a discussion of our results
in Section \ref{con}.

\section{Canonical analysis}

The first step in trying to understand the effects of the
topological terms added to the Holst action is to perform the
canonical analysis of our action. In order to do this it will be
convenient to write the topological terms in (\ref{action})
explicitly as exterior derivatives of $3$-forms, namely
\begin{widetext}
\begin{eqnarray}
  S[e,\omega] & = & \int \overbrace{ \alpha_1 \star \left( e^I \wedge
  e^J \right) \wedge F_{J I}(\omega) + \alpha_2 e^I \wedge e^J \wedge
  F_{I J}(\omega)}^{Holst} \nonumber \\ &+& (\alpha_3-i \alpha_4)d\sL_{CS}(\omega^{ASD})+
(\alpha_3+i \alpha_4)d\sL_{CS}(\omega^{SD})+ \alpha_5
d(d_{\omega}e^I \wedge e_I),
  \label{LHPE}
\end{eqnarray}
\end{widetext}
where $\omega^{SD}=i(\star\omega)^{SD}=\omega-i(\star\omega)$ and
$\omega^{ASD}=i(\star\omega)^{ASD}=\omega+i(\star\omega)$ are the
self-dual and anti-self dual parts of the Lorentz connection
$\omega$ respectively,  and \be \sL_{CS}(\omega)=\omega_{IJ}\wedge
d\omega^{IJ}+\frac{2}{3} \omega_{IJ}\wedge [\omega\wedge
\omega]^{IJ} \ee is the Chern-Simons Lagrangian density. Despite of
the presence of complex variables in the above expression of the
action, the action principle is manifestly real as the strict
equality with (\ref{action}) holds. In performing the canonical
analysis of our theory we will use both (\ref{action}) as well as
(\ref{LHPE}) according to convenience.

As mentioned above, the addition of boundary terms to the action
principle induces canonical transformations in the phase space
formulation. Notice that the terms added to Holst's formulation are
the most general total differentials that one can write using the
fields $e$ and $\omega$ without breaking local Lorentz invariance
and diffeomorphism invariance. Therefore, studying the canonical
structure behind (\ref{action}) amounts for studying the most
general set of possible canonical transformations compatible with
the field content and symmetries of the action.

We assume that the spacetime manifold has topology
${\sM}=\Sigma\times\R$, with $\Sigma$ compact. In order to perform
the Hamiltonian formulation we start by doing the customary $(3+1)$
decomposition consisting of choosing an arbitrary foliation of
spacetime in terms of the level hyper-surfaces of a global time
function $t$. The hyper-surfaces $t={\rm constant}$ will be denoted
$\Sigma$ as well. We denote $n^a$ the normal to the foliation. The
arbitrariness in the choice of foliation is encoded in the lapse
scalar $N$ and the shift vector (tangent to the foliation) $N^{a}$
which imply that the time vector $t^a$ (defined by $t^a(t)=1$) takes
the form $t^a=N^a+ N n^a$. This implies that the following equation
for the projection of the tetrad in the $t^a$ direction:
\begin{eqnarray}
  e_t ^I & = & N n^I + e^I_a N^a, \nonumber
\end{eqnarray}
where $n^I\equiv n^a e_a^I$. With these definitions the Holst
action (for the moment we are ignoring the topological terms, i.e.
taking $\alpha_3=\alpha_4=\alpha_5=0$) takes the simple form:
\begin{widetext}
\begin{eqnarray}
  \sL_H = 2 \epsilon^{a b c}  e_{a }^I e_{b }^J\, q_{IJKL} F^{KL}_{tc}
  +2N \,\epsilon^{a b c}  n^I e_{a }^J\, q_{IJKL} F^{KL}_{bc}+
  2N^d \, \epsilon^{a b c}  e_d^I e_{a }^J\, q_{IJKL}
  F^{KL}_{bc},\label{holsty}
\end{eqnarray}
\end{widetext}
where $q_{IJKL}=\alpha_1 \epsilon_{IJKL}+\alpha_2 \eta_{IJKL}$
denoting by $\eta_{IJKL}=\eta_{[I|K|}\eta_{J]L}$ the invariant
metric in the Lie algebra of the Lorentz group. If we define
$\Pi^a_{KL}=2\epsilon^{a b c}  e_{b }^I e_{c }^J\, q_{IJKL}$ then
the previous action takes the form \begin{eqnarray}
  && \nonumber \sL_H = \Pi^a_{I} \dot e^{I}_a+\Pi^a_{IJ} \dot \omega^{IJ}_a+ N^{IJ} D_a\Pi^a_{IJ}+
  2N \,\epsilon^{a b c}  n^I e_{a }^J\, q_{IJKL} F^{KL}_{bc}+N^a \Pi^{b}_{IJ}
  F^{IJ}_{ba}\\ &&\ \ \ \ \ \ \ \ \ \ \ \ \ \ \ \ \ \ \ \ \ \ \ \
  \ \ \ \ \ \ \ \ \ \ \ \ \ \ \ \ \ \ \ \ \ \ \ \ \ \
  \ \
  +{}^{\va (1)}\lambda_a^I \Pi^a_I+ {}^{\va (2)}\lambda_a^{KL}(\Pi^a_{KL}-2\epsilon^{a b c}  e_{b }^I e_{c }^J\, q_{IJKL}),
\label{aaa}\end{eqnarray} where $N^{IJ}\equiv\omega_t^{IJ}$, $N$,
$N^a$, ${}^{\va (1)}\lambda_a^I$, and ${}^{\va (2)}\lambda_a^{KL}$
are Lagrange multipliers imposing the primary constraints of the
Holst action. On the other hand $\Pi^a_{IJ}$ and $\Pi^a_I$ denote
the momentum conjugate to $\omega_a^{IJ}$ and $e_a^I$
respectively. Therefore,
the primary constraints are \ba && \Pi^a_I\approx 0 \label{seven} \\
&& \Pi^a_{KL}-2\epsilon^{a b c}  e_{b}^I e_{c }^J\, q_{IJKL}\approx 0\\
&& \mbox{Lorentz-Gauss law}\ \ \  D_a\Pi^a_{IJ}=2D_a(\epsilon^{a b c}  e_{b}^I e_{c }^J\, q_{IJKL})\approx 0 \\
&& \mbox{Vector constraint}\ \ \  \epsilon^{a b c}  e_d^I e_{a }^J\, q_{IJKL}
  F^{KL}_{bc}\approx 0 \\
&& \mbox{Scalar constraint}\ \ \ \epsilon^{a b c}  n^I e_{a }^J\,
q_{IJKL} F^{KL}_{bc}\approx 0.
 \ea
A simple look at the list of primary constraints tell us that
there will be secondary constraints when we require the primary to
be preserved by the Hamiltonian evolution. However, before
continuing and completing the analysis it will be convenient to
treat the general case including the topological terms.

\subsection{The Holst action plus topological terms}

Including the topological terms is straightforward at this level.
According to (\ref{LHPE}), and using that $\partial\Sigma=0$, the
Lagrangian (\ref{holsty}) is modified by the addition of the total
time derivative, namely \be\sL=\sL_H+(\alpha_3-i\alpha_4) \partial_t
\sL_{CS}(\omega^{ASD})+(\alpha_3+i\alpha_4) \partial_t
\sL_{CS}(\omega^{SD})+\alpha_5 \partial_t (d_{\omega}e^I\wedge e_I),
\label{toplus}\ee which, using that $\partial_t
\sL_{CS}(\omega)=2B^a_{IJ}(\omega)\dot \omega_{a}^{IJ}$ where
$B^a_{IJ}=\epsilon^{abc}F_{bc}(\omega)$, implies that the conjugate
momenta $\Pi^a_{IJ}$ and $\Pi^a_I$ receive additional contributions
of the form $4\alpha_3B^a_{IJ}+ 4\alpha_4 \epsilon_{IJ}^{\ \
KL}B^a_{KL}+\alpha_5 \epsilon^{abc} (e_{b})_{[I}(e_{b})_{J]}$ and
$2\alpha_5 \epsilon^{abc}d_{\omega}e^I_{bc}$ respectively. Notice
that, due to $\partial \Sigma=0$, the addition of the topological
terms only affects the kinetic term of the Holst action. More
precisely if we define the real functional  \be
W(\omega_a^{IJ},e_a^I)=\int_{\Sigma}(\alpha_3-i\alpha_4)
\sL_{CS}(\omega^{ASD})+(\alpha_3+i\alpha_4)
\sL_{CS}(\omega^{SD})+\alpha_5 (d_{\omega}e^I\wedge e_I),
\label{toplusy}\ee the new constraints become \ba && \nonumber
\Pi^a_I-\frac{\delta W}{\delta {e_{a}^{I}}}=
 \\ && =
\Pi^a_I-2\alpha_5 \epsilon^{abc}(d_{\omega}e)_{bc I}\approx 0 \label{thirdteen} \\
&& \nonumber \Pi^a_{IJ}-2\epsilon^{a b c}  e_{b}^K e_{c }^L\,
q_{IJKL}-\frac{\delta W}{\delta {\omega_{a}^{IJ}}}=
 \\ && = \Pi^a_{KL}-2\epsilon^{a b c}  e_{b}^I e_{c }^J\,
q_{IJKL}-\alpha_5 \epsilon^{abc}
(e_{b})_{[K}(e_{b})_{L]}-4\alpha_3B^a_{KL}- 4\alpha_4
\epsilon_{KL}^{\ \ IJ}B^a_{IJ}\approx 0\\\nonumber
\\
&& \mbox{Lorentz-Gauss law}\ \ \  D_a\Pi^a_{IJ}=2D_a(\epsilon^{a b c}  e_{b}^I e_{c }^J\, q_{IJKL})\approx 0 \label{16}\\
&& \mbox{Vector constraint}\ \ \  \epsilon^{a b c}  e_d^I e_{a }^J\, q_{IJKL}
  F^{KL}_{bc}\approx 0 \\
&& \mbox{Scalar constraint}\ \ \ \epsilon^{a b c}  n^I e_{a }^J\,
q_{IJKL} F^{KL}_{bc}\approx 0.
 \ea
Notice that the addition of the topological invariants, being just
boundary terms, modify only the constraints defining the momenta.
The vector and scalar constraints remain the same as they do not
depend on momentum variables at this stage ($e_a^I$ is here
considered a configuration variable). The Gauss law (\ref{16}) does depend on
the momenta (as written as in the l.h.s.); however, it
also remains  unchanged (as in the r.h.s.) due to the Bianchi identity implying
$D_aB^a_{IJ}=0$.

\subsection{The time gauge: reducing $SO(3,1)$ to $SO(3)$}

Let us now introduce the standard gauge condition that reduces the
Lorentz gauge symmetry to an $SO(3)$ gauge symmetry. The gauge
condition is often called the time-gauge condition. It corresponds
to the requirement that the zeroth element of the tetrad coincide
with the co-normal to $\Sigma$, namely $n_I e^I_{\mu}=n_{\mu}$. This
implies the phase-space additional gauge-fixing constraint
\be\label{timegauge} n_Ie_a^I\approx0,\ee which now must be added to
the list of primary constraint above. The previous gauge fixing
condition is necessary to recover the compact gauge group connection
variables that are used in LQG. One can of course complete the
Hamiltonian formulation without breaking the local Lorentz
invariance. However, the price to be paid is a non trivial Dirac
bracket between the components the Lorentz connection
$\omega_{a}^{IJ}$ \cite{sergey} precluding the existence of a
connection representation in the quantum theory. A proposal for
quantizing the non-commutative connection can be found in
\cite{sergey1}.

The condition (\ref{timegauge}) is second class with respect to the
projection of equation (\ref{seven}) in the $n^I$ internal
direction. In other words the requirement that the gauge
(\ref{timegauge}) is preserved in time fixes the Lagrange
multipliers ${}^{\va (1)}\lambda^a_I$ in (\ref{aaa}). This means
that we can impose $n^I\Pi_I^a=0$ and $n_Ie_a^I=0$ strongly. From
now on we will take $n^I=(1,0,0,0)$ and denote with lower case Latin
alphabet letters the space-like internal directions. The new
restricted dynamical system is described by
\ba && \Pi^a_i-2\alpha_5 \epsilon^{abc}(d_{\omega}e)_{bc i} \approx 0 \label{se} \\
&& \frac{1}{2}\epsilon_{k}^{\ ij}\Pi^a_{ij}\approx
(2\alpha_2-\alpha_5) \epsilon^{a b c} e_{b}^i e_{c }^j\epsilon_{ijk}
+4\alpha_3\epsilon_{k}^{\ ij} B^a_{ij}- 16 \alpha_4 B^a_{k0} \label{ku}\\
&& \Pi^a_{k0}\approx 2\alpha_1 \epsilon^{a b c}  e_{b}^i e_{c
}^j\, \epsilon_{ijk}+4\alpha_3B^a_{k0}- 4\alpha_4 \epsilon_{k}^{\
\
lm}B^a_{lm}\label{ki}\\
&& \label{we}\mbox{Lorentz-Gauss law}\ \ \ \left\{\begin{array}{ccc} \epsilon_{m l k} E^{a l} \hat{K}^k_a &\approx& 0, \\
\partial_d E^{d k} + \epsilon_{i j k} \hat{\Gamma}^i_d E^{d j} &\approx& 0,\end{array}\right.\\
&& \mbox{Vector constraint}\ \ \  \Pi^{b}_{ij} F^{ij}_{ba}+2\Pi^{b}_{i0} F^{i0}_{ba}\approx 0 \\
&& \mbox{Scalar constraint}\ \ \ \epsilon^{a b c}  e_{a }^i\,
q_{0ijk} F^{jk}_{bc}+2 \epsilon^{a b c}  e_{a }^i\, q_{0i0k}
F^{0k}_{bc}\approx 0,
 \ea
 where we have used the following definitions
\be E^a_i\equiv\frac{1}{2}\epsilon^{abc}e^j_ae^k_b\epsilon_{ijk}, \
\ \ \ \hat K_a^i\equiv \omega_a^{0i}\ \ \ \ \hat
\Gamma_{a}^i\equiv\frac{1}{2} \epsilon^i_{\ \ jk}
 \omega^{jk}_a \label{triad}\ee and the Bianchi identity to write the Gauss
law constraints (\ref{we}). We will see in a moment that the
previous variables are
 indeed the extrinsic curvature component and the Levi-Civita spin
 connection respectively, which justifies the notation.
Equations (\ref{ku}) and (\ref{ki}) can be combined in a way to
simplify the dependence on the triad $e_a^i$: notice that the triad
dependence is the same in both equations. Therefore, one can
introduce new variables \ba \nonumber {\Ppm}^a_i&\equiv&
\frac{1}{4}\epsilon_{i}^{\ jk}
\Pi^{a}_{jk}\pm \frac{2\alpha_2+\alpha_5}{4\alpha_1}\Pi^a_{i0} \\
&=&\frac{1}{4}\epsilon_{i}^{\ jk} \Pi^{a}_{jk}\pm
\frac{1}{2\gamma}\Pi^a_{i0}, \ea where we introduced the definition
$\gamma\equiv\frac{2\alpha_1}{2\alpha_2-\alpha_5}$. The previous new
momenta are the conjugate of new $SO(3)$ connections
\begin{equation}
    ^{\va \pm}\!\omega_{d l} = \pm {\gamma}\omega_{d l0} + \frac{1}{2} \epsilon_{l}^{\ m
    n} \omega_{d m n},
\end{equation}
and we recognize $\gamma$ as the Immirzi parameter at this stage. In
the time gauge, one can write the functional
$W(\omega_a^{IJ},e_a^I)$ defined in (\ref{toplusy}) as a functional
as $W(\omega_a^{IJ},e_a^I)=W_0(\op_a^i,\om_a^i)+\alpha_5
\epsilon^{abc}(d_{\omega} e)^i_{ab}e_{ci}$
 where $W_0(\op_a^i,\om_a^i)$  is simply the value of $W(\omega_a^{IJ},e_a^I)$
 for $\alpha _5=0$.
 Using the new
variables the constraints become
\ba &&  \mathbbm{(I)}^a_i\equiv \Pi^a_i-2\alpha_5 \epsilon^{abc}(d_{\omega}e)_{bc i} \approx 0 \label{sel} \\
&& \label{magui}  \mathbbm{(II)}^a_k\equiv \Pp^a_k - 2
\frac{\alpha_1}{\gamma} \epsilon^{a b c} e_{b}^i e_{c
}^j\epsilon_{ijk}
-\frac{\delta W_0}{\delta \op_a^i}\label{kul}\\
&&  \mathbbm{(III)}^a_k\equiv \Pm^a_k-\frac{\delta W_0}{\delta \om_a^i}\label{kul}
\label{kil}\\ \nonumber\\
&& \mbox{Boosts constraint} \ \ \ B^k\equiv\partial_d E^{d k} - \epsilon_{i j k} \hat{\Gamma}^i_d E^{d j} \approx 0 \Longrightarrow B^k=-\epsilon_{ijk} E^{ai}(\hat\Gamma_a^j-\Gamma_a^j)\approx 0\label{ggg}\\
&& \label{gausso3}\mbox{$SO(3)$ Gauss law}\ \ \  G_m \equiv \epsilon_{m l k} E^{a l} \hat{K}^k_a \approx 0 \\
&& \mbox{Vector constraint}\ \ \  V_a\equiv \Pi^{b}_{ij} F^{ij}_{ba}+2\Pi^{b}_{i0} F^{i0}_{ba}\approx 0 \\
&& \mbox{Scalar constraint}\ \ \ S\equiv \epsilon^{a b c}  e_{a }^i\,
q_{0ijk} F^{jk}_{bc}+2 \epsilon^{a b c}  e_{a }^i\, q_{0i0k}
F^{0k}_{bc}\approx 0,
 \ea
where \ba \hat{K}_a^i = \frac{1}{2\gamma} ({ \op_{a}^i }-\om_{a}^i
) \ \ \ \ \   \hat{\Gamma}_a^i &=& \frac{1}{2}({\om_{a}^i +
\op_{a}^i }). \label{OffShellGamma}
\end{eqnarray}
Notice that we have re-written the boost part of the Lorentzian
Gauss law---which we should expect to be second class due to the
time gauge condition (\ref{timegauge})---in terms of the spin
connection $\Gamma_a^i$, i.e., the solution of Cartan's first
structure equation
\begin{eqnarray}
\partial_{[a} e^{k}_{b]} - \epsilon_{\ \ i j}^{ k}
\Gamma^i_{[a} e^{j}_{b]}= 0.
\end{eqnarray}
Indeed it will be convenient to introduce the quantity
\be
\mathbbm{(IV)}_a^i\equiv \hat \Gamma_a^i-\Gamma_a^i.
\label{4}
\ee
We will explicitly show in what follows how three components of the boost part of the
Lorentzian Gauss law plus six secondary constraints (not derived yet) imply
$\mathbbm{(IV)}_a^i\approx 0$ which will be shown to be second class.

By setting $W=0$ one recovers the primary constraints of Holst
\cite{holst}. Notice in addition that, as mentioned above, only the
first three constraints in the previous list are modified by the
addition of the Pontrjagin, Euler and Nieh-Yan invariants to the
Holst action. The modification is very simple: if we take
$\{\op_a^i,\om_a^i, e_a^i\}$ as configuration variables then, the
Holst momenta are shifted according to $p\rightarrow p+\{p ,W({}^+
\omega_a^i,\om_a^k,e_a^i)\}$, where $p$ denotes $\Pp^a_i,
\Pm^a_{i}$, and $\Pi^a_i$. The modification introduced by the
topological invariants is just a canonical transformation generated
by $W({}^+ \omega_a^i,\om_a^k,e_a^i)$. For that reason, the
constraint algebra is not affected by the topological terms.
Therefore, the consistency conditions (secondary constraints) that
follow from requiring that primary constraints are preserved by the
total Hamiltonian remain unchanged. This also hold for the
classification between first class and second class constraints.
Collecting all primary constraints the Hamiltonian becomes
\begin{eqnarray}\label{39}
  {H}= \int_{\Sigma} N S + N^a V_a + N^k G_k + \lambda^k
  B_k + \lambda^{\va (1)}_{d l} (\mathbbm{I})^{d l} + \lambda^{\va (2)}_{d l} (\mathbbm{II})^{d l} + \lambda^{\va (3)}_{d l} (\mathbbm{III})^{d l},
\end{eqnarray}
At this point one needs to look for potential secondary constraint by requiring that the constraint surface be
preserved by the time evolution defined by the previous Hamiltonian. This leads to the following
consistency conditions:
\begin{eqnarray}
  && 0\approx \left\{ \mathbbm{(I)}^a_i, {H}_T
  \right\}=\frac{4\alpha_1}{\gamma}\epsilon^{abc}\lambda_b^{{\va
  (2)}j}e_c^k\epsilon_{ijk}+ \mbox{additional terms} \label{cons1}\\
  && 0\approx \left\{ \mathbbm{(II)}^a_i, {H}_T
  \right\}=\frac{1}{2}\lambda^{j} E^{a k}\epsilon_{ijk}
  -(\alpha_5+2(2\alpha_2-\alpha_5)) \epsilon_{ijk}\epsilon^{abc}e_b^j\lambda^{{\va(1)}k}_c-\frac{\delta H_0}{\delta^{+}\omega_a^k
  }  \label{cons2}\\
  &&  0\approx \left\{\mathbbm{(III)}^a_i, {H}_T
  \right\}=\frac{1}{2}\lambda^{j} E^{a k}\epsilon_{ijk}-\alpha_5 \epsilon_{ijk}\epsilon^{abc}e_b^j\lambda^{{\va(1)}k}_c-\frac{\delta H_0}{\delta^{-}\omega_a^k
  } \label{cons3}\\
  && 0\approx \left\{ B_i, {H}_T
  \right\}=-\frac{1}{2}\epsilon_{ijk}\lambda_b^{{\va
  (3)}j}E^{b k}+ \mbox{additional terms} \label{cons4},\ea
where $H_0\equiv\int_{\Sigma} N S + N^a V_a +N^kG_k$ is a linear
combination of the scalar, vector and $SO(3)$ Gauss constraints.
Equations (\ref{cons1}) and (\ref{cons2}) completely determine the
Lagrange multipliers $\lambda_b^{{\va
  (2)}j}$ and $\lambda_b^{{\va
  (1)}j}$ respectively. Equation (\ref{cons4}) determines the antisymmetric part of
  $\lambda_{ij}^{{\va
  (3)}}\equiv \lambda_{bi}^{{\va
  (2)}}E^b_j$, i.e., it fixes three out of the nine components of the Lagrange multiplier
  $\lambda_b^{{\va
  (3)}j}$. Hence, there are no secondary constraints
  arising from these equations.

  Equation (\ref{cons3}) leads to
  secondary constraints. To see this one has to replace in
  (\ref{cons3}) the solution for $\lambda^{{\va (1)}i}_a$
  obtained from (\ref{cons2}). Consider the quantity
\be C_{ij}\equiv(e_a)_j \left\{ \mathbbm{(III)}^a_i, {H}_T
\right\}. \ee Then it is easy to see that the three conditions
$C_{[ij]}=0$ can be used to fix the Lagrange multipliers
$\lambda^i$ while the remaining six conditions $C_{(ij)}=0$ are
proportional to
\begin{eqnarray}
 C_{(ij)}=(e_a)_{(j} \left\{ \mathbbm{(III)}^a_{i)}, {H}_T
\right\}\propto \epsilon^{a b d} E_{d (m} d^{ (\hat \Gamma)}_a
e_{b l)} \approx 0, \label{eq6}
\end{eqnarray}
where $d^{\hat \Gamma}$  is the exterior covariant differential
computed with $\hat \Gamma$. Notice now that the six independent
above constraints over $\hat{\Gamma}$ can be combined with the
three boost constraints $B^i$ in (\ref{ggg}) into  the nine
component constraint (\ref{4}), namely
\begin{eqnarray}
B^{i}=0 \ \ \ \mbox{ in addition to}\ \ \ C_{(ij)}=0 \ \ \
\Leftrightarrow\ \ \  \mathbbm{(IV)}_a^i\equiv \hat
\Gamma_a^i-\Gamma_a^i=0. \label{PsiConstr}
\end{eqnarray}
We can therefore rearrange the total Hamiltonian in the more
convenient form \be {H}_T = \int_{\Sigma} N S + N^a V_a + N^k G_k +
\lambda^{\va (1)}_{d l} (\mathbbm{I})^{d l} + \lambda^{\va (2)}_{d
l} (\mathbbm{II})^{d l} + \lambda^{\va (3)}_{d l} (\mathbbm{III})^{d
l}+ \lambda^{\va (4)d }_l (\mathbbm{IV})_{d}^{l}, \ee where instead
of adding the secondary constraint (\ref{eq6}) to the primary
constraints Hamiltonian,
 we have dropped the term $\lambda^k
  B_k$ from the integrand in (\ref{39}) and added the term $\lambda^{\va (4)d }_l (\mathbbm{IV})_{d}^{l}$
 in the previous expression of the total Hamiltonian.
 One can check
that the consistency conditions of the new set of constraints fix
the Lagrange multipliers $\lambda^{\va (\mu)}$ for $\mu=1,2,3,4$,
while the Lagrange multipliers $N,N^a$ and $N^k$ are left arbitrary.
From this one concludes that the $36$ constraints
$(\mathbbm{I})^a_i$, $(\mathbbm{II})^a_i$, $(\mathbbm{III})^a_i$,
and $(\mathbbm{IV})_a^i$ are second class constraint while the seven
remaining constraints (the scalar $S$, vector $V_a$ and Gauss $G_k$
constraints) are first class\footnote{Strictly speaking the scalar
$S$, vector $V_a$ and Gauss $G_k$ constraints are not first class as
written here. In order to make them into first class constraints one
would need to add to them appropriate linear combinations of the
second class constraints. Nevertheless, the fact that $N,N^a$ and
$N^k$ are not fixed by the equations of motion implies the existence
of 7 first class constraints and that these coincide with the scalar
$S$, vector $V_a$ and Gauss $G_k$ constraints once the second class
constraints have been solved. To see this in a more general way,
suppose we have two sets of constraints, ${\phi_A}$ and ${\theta_A}$
such that
\begin{eqnarray*}
\{ \phi_A, \phi_B \} &=& 0, \\
\{ \phi_A, \theta_B \} &=& M_{A B}, \\
\{ \theta_A, \theta_B \} &=& \Delta_{A B},
\end{eqnarray*}
where $\Delta$ is an invertible matrix and $M$ not. We can `decouple' the two sets of constraints by the redefinition:
\[
\phi_A \rightarrow \tilde{\phi}_A = \phi_A - M_{A C} (\Delta^{-1})^{C D}\theta_D,
\]
the algebra becomes:
\begin{eqnarray*}
\{ \tilde{\phi}_A, \tilde{\phi}_B \} &=& \{\phi_A - M_{A C} (\Delta^{-1})^{C D}\theta_D,\phi_B - M_{B C} (\Delta^{-1})^{C D}\theta_D\}\approx 0, \\
\{ \tilde{\phi}_A, \theta_B \} &=& -\{M_{A C} (\Delta^{-1})^{C D},\theta_B\}\theta_D \approx 0, \\
\{ \theta_A, \theta_B \} &=& \Delta_{A B},
\end{eqnarray*}
this is completely equivalent to solve first the Dirac brackets for the ${\theta_A}$ sector and then, recomputing the remaining algebra for the ${\phi_A}$ sector.}.
There are 27 configuration variables $e_a^i$, $\op_a^i$, and
$\om_a^i$; therefore, the counting of degrees of freedom yields
the expected two degrees of freedom of gravity.

\section{Construction of the Dirac bracket}\label{dirac}

The analysis up to this point follows the same logical line as in
the Holst's case, by the simple fact that the addition of a surface
term to the action does not change the Poisson brackets among the
constraints. However, contrary to the Holst case, we cannot
explicitly solve the constraint $(\mathbbm{II})^a_i$. The reason is
the presence of the curvature tensor in the magnetic field
contributions to (\ref{magui}) which prevents one from eliminating
the densitized triad as a function of the connection and its
momenta: the dependence on the densitized triad on is quite
complicated due to the non polynomial character of the spin
connection $\Gamma_a^i(E)$. Thus, in order to complete the canonical
analysis of the general action, we need to explicitly construct the
Dirac brackets for the second class constraints $(\mathbbm{I})^a_i$,
$(\mathbbm{II})^a_i$, $(\mathbbm{III})^a_i$, and
$(\mathbbm{IV})_a^i$. Before computing the constraint algebra it
will be convenient to replace the constraint $(\mathbbm{I})^a_i$.
The constraint algebra is
\begin{eqnarray}
&& \left\{ (\mathbbm{I})^a_i,(\mathbbm{I})^b_j \right\}  =  0, \\
&& \left\{ (\mathbbm{I})^a_i,(\mathbbm{II})^b_j \right\}  =
\left(4\alpha_2-\alpha_5\right) \epsilon^{abc}\epsilon_{ijk}e_c^k \delta^3 (x, y),\\
&& \left\{ (\mathbbm{I})^a_i,(\mathbbm{III})^b_j \right\}  =  \alpha_5 \epsilon^{abc}\epsilon_{ijk}e_c^k \delta^3 (x, y), \\
&& \left\{ (\mathbbm{I})^a_i,(\mathbbm{IV})_b^j \right\}  =
\frac{\delta \Gamma^{j}_{b} ( y )}{\delta e_a^i ( x )},\\
&& \left\{ (\mathbbm{II})^a_i,(\mathbbm{II})^b_j \right\}  =  0,\\
&& \left\{ (\mathbbm{II})^a_i,(\mathbbm{III})^b_j \right\}  =
0,\\
&& \left\{ (\mathbbm{II})^a_i,(\mathbbm{IV})_b^j \right\}  =-
\frac{1}{2}\delta^{b}_a \delta^j_i \delta^3 (x, y),\\ &&
\left\{
(\mathbbm{III})^a_i,(\mathbbm{III})^b_j \right\}  = 0,\\
&& \left\{(\mathbbm{III})^a_i,(\mathbbm{IV})_b^j \right\}  =
-\frac{1}{2} \delta^{b}_a \delta^j_i \delta^3 (x, y), \\ &&
\left\{(\mathbbm{IV})^a_i,(\mathbbm{IV})^b_j \right\}=0. \ea

We construct the Dirac matrix, which we represent symbolically as
\[ \mathbbm{M} = \left(\begin{array}{cccc}
     0 & (4\alpha_2-\alpha_5) A^{ab}_{ij}(x, y) & \alpha_5  A^{ab}_{ij}(x, y) & B^{bk}_{al}(x, y)\\
     - (4\alpha_2-\alpha_5) A^{ab}_{ij}(x, y) & 0 & 0 & -\frac{1}{2}I^{bk}_{al}(x, y)\\ -\alpha_5  A^{ab}_{ij}(x, y) & 0 & 0 & -\frac{1}{2} I^{bk}_{al}(x, y)\\
     -B^{bk}_{al}(x, y) & \frac{1}{2}I^{bk}_{al}(x, y) & \frac{1}{2}I^{bk}_{al}(x, y) & 0
   \end{array}\right), \]
where
\[
I^{bk}_{al}(x, y) = \delta^b_a \delta^k_l \delta^3 (x, y),
\]
\[
A^{ab}_{ij}(x, y) = \epsilon^{abc}\epsilon_{ijk}e_c^k \delta^3 (x, y),
\]
and
\[
B^{bk}_{al}(x, y) \equiv \frac{\delta \Gamma^k_a ( x )}{\delta e_b^l ( y
)}.
\]
The inverse of this matrix is:
\[
\mathbbm{M}^{- 1} =\frac{\gamma}{4\alpha_1} \left(\begin{array}{cccc}
     0 & -(A^{-1})_{ab}^{ij}(x, y) & (A^{-1})_{ab}^{ij}(x, y) & 0\\
    (A^{-1})_{ab}^{ij}(x, y) & 0 & -2 (A^{-1}\cdot B)^{kl}_{ab}(x, y) & -2\alpha_5  I^{bk}_{al}(x, y)\\
    - (A^{-1})_{ab}^{ij}(x, y) &2 (A^{-1}\cdot B)^{kl}_{ab}(x, y) & 0 & 2(4\alpha_2-\alpha_5) I^{bk}_{al}(x, y) \\
    0& 2\alpha_5  I^{bk}_{al}(x, y) & -2(4\alpha_2-\alpha_5) I^{bk}_{al}(x, y) & 0
   \end{array}\right) ,
\]
where the dot in the above expression involves the appropriate index
contraction and integration over $\Sigma$, explicitly: \be
(A^{-1}\cdot B)^{kl}_{ab}(x, y)\equiv \int dz (A^{-1})_{ac}^{km}(x,
z) B^{cl}_{bm}(z, y). \ee The only explicit inversion that one needs
is that of the tensor density $A^{ab}_{ij}(x, y)$. It is
straightforward to show that the inverse is given by
\be(A^{-1})_{ab}^{ij}(x, y)=\frac{\delta e_a^i(x)}{\delta
E^b_j(y)},\ee which can be computed explicitly using that
$e_a^i=\frac{1}{2}\epsilon_{abc}\epsilon^{ijk}
E^b_jE^c_k/(\sqrt{{\rm det}(E)})$ as implied by eq. (\ref{triad}).
Notice also that the previous equation implies \be(A^{-1}\cdot
B)^{kl}_{ab}(x, y)= \frac{\delta \Gamma^k_a ( x )}{\delta
E^b_l(y)}.\ee Thus, the full Dirac bracket is given by {
\begin{eqnarray} \left\{f , g \right\}_D & =& \left\{ f, g \right\} -\frac{\gamma}{4\alpha_1}
\int  \left[ - \{ f, (\mathbbm{I})^a_i(x)
\}
 \frac{\delta e^i_a ( x )}{\delta
E^b_j(y)} \{ (\mathbbm{II})^b_j(y), g \} + \{ f, (\mathbbm{I})^a_i(x) \}
 \frac{\delta e^i_a ( x )}{\delta
E^b_j(y)} \{ (\mathbbm{III})^b_j(y), g \}\nonumber \right. \\ &+&\left.
 \{ f, (\mathbbm{II})^a_i(x) \}
 \frac{\delta e^i_a ( x )}{\delta
E^b_j(y)} \{ (\mathbbm{I})^b_j(y), g \}-2 \{ f, (\mathbbm{II})^a_i(x) \}
 \frac{\delta \Gamma^i_a ( x )}{\delta
E^b_j(y)} \{ (\mathbbm{III})^b_j(y), g \}\nonumber \right. \\ &-&\left. 2\alpha_5 \{ f,
(\mathbbm{II})^a_i(x) \}
 \{ (\mathbbm{IV})_a^i(y), g \} -\{ f, (\mathbbm{III})^a_i(x) \}
 \frac{\delta e^i_a ( x )}{\delta
E^b_j(y)} \{ (\mathbbm{I})^b_j(y), g \} \right. \nonumber \\ & + &
\left.  2 \{ f, (\mathbbm{III})^a_i(x) \}
 \frac{\delta \Gamma^i_a ( x )}{\delta
E^b_j(y)} \{ (\mathbbm{II})^b_j(y), g \} + 2(4\alpha_2-\alpha_5) \{ f,
(\mathbbm{III})^a_i(x) \}
 \{ (\mathbbm{IV})_a^i(y), g \}\right. \nonumber \\ & + &\left. 2\alpha_5 \{ f,
(\mathbbm{IV})^a_i(x) \}
 \{ (\mathbbm{I})^a_i(y), g \} - 2(4\alpha_2-\alpha_5) \{ f,
(\mathbbm{IV})^a_i(x) \}
 \{ (\mathbbm{III})_a^i(y), g \} \frac{{}^{{}^{}}}{{}_{{}_{}}} \right] d^3 x d^3 y\label{DIRACBR}
.\end{eqnarray} From (\ref{DIRACBR}) we obtain the new commutation
relations:
\begin{eqnarray} &&\nonumber
 \left\{ \op_a^i ( x ), e_{b}^{j} ( y ) \right\}_D =
  \frac{\gamma}{4 \alpha_1} \frac{\delta e^i_a ( x )}{\delta
E^b_j(y)} , \label{DBAE} \\\nonumber
  &&
 \left\{ \op_a^i ( x ),\op_{b}^{j} (y) \right\}_D = 0,  \\ \nonumber
&&
\left\{ \op_a^i ( x ), \om_b^j ( y ) \right\}_D = \frac{\gamma}{2\alpha_1}\frac{ \delta \Gamma_b^j(y)}{ \delta E^a_k ( x )},\\ \nonumber
&&
\left\{ e_{a}^i ( x ), e_{b}^{j} ( y ) \right\}_D= 0, \\ \nonumber
  &&
\left\{ e_a^k ( x ),\Pp^{b}_{j} ( y ) \right\}_D =-\frac{\gamma}{4
  \alpha_1} \int \frac{ \delta e_d^m(z)}{ \delta E^a_i ( x )}
  \frac{\delta}{\delta\op_d^m ( z )}\left[ \frac{\delta W_0}{\delta \op_{b}^{j} ( y )}
  -\frac{\delta W_0}{\delta \om_{b}^{j} ( y )}
  \right] dz, \label{BREP} \\ \nonumber
  &&
\left\{ e_a^k ( x ),\Pm^{b}_{j} ( y ) \right\}_D =
-\frac{\gamma}{4
  \alpha_1} \int \frac{ \delta e_d^m(z)}{ \delta E^a_i ( x )}
  \frac{\delta}{\delta\om_d^m ( z )}\left[ \frac{\delta W_0}{\delta \op_{b}^{j} ( y )}
  -\frac{\delta W_0}{\delta \om_{b}^{j} ( y )}
  \right] dz, \label{BREP}\\&&\nonumber
\left\{\Pp^a_i ( x ),\Pp^b_j ( y ) \right\}_D =
  -\frac{\gamma}{2\alpha_1}\,  \int\left[\frac{ \delta^2
  W_0 }{\delta \op_{a}^{i} ( y )\delta \op_c^{k}(z)}
\frac{ \delta \Gamma_c^k(z)}{ \delta E^d_m ( w )}
 \frac{ \delta^2
  W_0 }{\delta \om_{d}^{m} ( w )\delta \op_b^{j}(y)}- ({}_a^i) \leftrightarrow ({}_b^j)\right] dzdw, \\ \nonumber
 &&
\left\{\Pm^a_i ( x ),\Pm^b_j ( y ) \right\}_D =
  -\frac{\gamma}{2\alpha_1}\,  \int\left[\frac{ \delta^2
  W_0 }{\delta \om_{a}^{i} ( y )\delta \op_c^{k}(z)}
\frac{ \delta \Gamma_c^k(z)}{ \delta E^d_m ( w )}
 \frac{ \delta^2
  W_0 }{\delta \om_{d}^{m} ( w )\delta \om_b^{j}(y)}- ({}_a^i) \leftrightarrow ({}_b^j)\right] dzdw, \\ \nonumber
&& \left\{ \om_a^i( x ), \Pm^{b}_{j} ( y )
  \right\}_D =  \frac{\gamma}{2 \alpha_1}
  \int\left[ \frac{ \delta \Gamma_d^m(z)}{ \delta E^a_i ( x )} \frac{ \delta^2
  W_0 }{\delta \om_{b}^{j} ( y )  \delta \op_d^m ( z)} \right] dz, \label{BRAPy}\\
&& \left\{ \op_a^i( x ), \Pp^{b}_{j} ( y )
  \right\}_D =\delta^b_a\delta^i_j \delta (x,y) - \frac{\gamma}{2 \alpha_1}
  \int\left[\frac{ \delta \Gamma_d^m(z)}{ \delta E^a_i ( x )} \frac{
  \delta^2
  W_0 }{\delta \op_{b}^{j} ( y ) \delta \om_d^m ( z)} \right] dz, \label{BRAP}
\end{eqnarray}
If from now on we use only the Dirac brackets, we are allowed to
eliminate all the second class constraints, in particular
(\ref{PsiConstr}). Thus, writing everything only as functions of
$\op_{d l}$ and $E^{d l}$ the scalar and vector constraints become
\begin{eqnarray}
  && S = 4 \alpha_1 e^{- 1} \epsilon^{i j k} E^b_j E^c_k \left[
  -\gamma^2 F_{b c i}(\op) + \left( \gamma^2 + 1 \right) F_{b c i}
  (\Gamma) \right], \\ && \nonumber V_d = - 4 \alpha_1 \left[ \gamma
  E^{c k} F_{d c k}(\op) + \left( \gamma^2 + 1 \right) K^k_d [^{+}
  \mathcal{D}_c E^c ]_k \right].\ea
 The expression of $\Pp$ as a function of $\op_{d l}$ and $E^{d l}$ is
\ba && \Pp^d_k = 4 \alpha_2
  E^d_k +\epsilon^{dab} \left[ ( 4\alpha_4 \gamma -4 \alpha_3 \gamma^2
  ) F_{abk}(\op) + \alpha_3( 1+\gamma^2 ) F_{abk}(\Gamma)
  \right. \nonumber \\ \nonumber && \ \ \ \ \ \ \ \ \ \ \ \ \ \ \ \ \
  \ \ \ \ \ \ \ \ \ \ \ \ \ \ \ \ \ \ \ \ \ \ \ \ \ \ \ \ \ \ \ \ \ \ \
  \ \ \ \ \left.  + ( 6 \alpha_3 + 2\alpha_4 \gamma - 4
  \frac{\alpha_4}{\gamma}) \epsilon_{k m n} K^m_a K^n_b \right].
\end{eqnarray}

\section{Which $W_0$ lead to canonical transformations?}\label{cano}

The Dirac algebra among the basic variables found
in the previous section for arbitrary $W_0$ is quite complicated.
In this section we investigate the possible choices of $W_0$ such
that $(4\alpha_1 E/\gamma, \op)\to (\Pp, \op)$ is a canonical transformation.

The necessary and sufficient condition that one needs to satisfy is
\be\label{uy} \int\left[\frac{ \delta \Gamma_d^m(z)}{ \delta E^a_i (
x )} \frac{
  \delta^2
  W_0 }{\delta \op_{b}^{j} ( y ) \delta \om_d^m ( z)} \right] dz= 0
\ee for all field configurations and for $W_0$ a functional of $\op$
and $\om$ respectively. This condition holds if and only if \be
 \frac{
  \delta^2
  W_0 }{\delta \op_{b}^{j} ( y ) \delta \om_d^m ( z)}= 0,
\label{clave}\ee whose solution is given by
$W_0[\op,\om]=W^+_0[\op]+ W^-_0[\om]$. We should point out that the
integral equation \be \int \frac{ \delta \Gamma_d^m(z)}{ \delta
E^a_i ( x )} V_m^d(z,y) dz= 0 \label{uyy}\ee admits non trivial
solutions. Recall that each $E^a_i$ gives a unique spin connections
  $\Gamma_a^i$; however, this relationship is not invertible. The reason is
  that $\Gamma_a^i(E)=\Gamma_a^i(\lambda E)$ for
  $\lambda=$constant. Therefore, only the scale invariant geometry ($E$ up
to a constant factor) can be recovered from $\Gamma_a^i$. This
implies a non trivial solution of equation (\ref{uyy}), for instance
 $V^a_i(x,y)=E^a_i(x) \Omega(y)$. Nevertheless, that solution depends explicitly on $E$ and
cannot be realized by derivatives of $W_0$. This in turn implies
that the canonical transformation takes the simple form \be
\Pp^a_i=4\frac{\alpha_1}{\gamma} E^a_i+ \frac{
  \delta
  W^+_0 }{\delta \op_{a}^{i}}. \label{equs}
\ee Recall that we started from the most general action principle for
general relativity in the tetrad first order formulation. Therefore,
in addition to the factorization property written above, the
generating function must derive from a particular combination of the
Pontrjagin, Euler and Nieh-Yan invariants (which are Lorentz
invariant). The general solution to these constraints is $\gamma=\pm
i$ (or equivalently $-2i\alpha_1=2\alpha_2-\alpha_5$) and $\alpha_3$
and $\alpha_4$ arbitrary. The canonical transformation (\ref{equs})
becomes in this case \be \Pp^a_i=\mp i4{\alpha_1}E^a_i+ 2
(\alpha_3\pm i \alpha_4) \epsilon^{abc} F_{bci}(\op). \label{equ}
\ee which corresponds to the one obtained in \cite{merced} for the
special case $\alpha_5=0$ and $\alpha_3=i \alpha_4$. For instance
when $\alpha_5=0$ the action becomes \ba S[e,\omega] & = & \int
\alpha_1 \left( e^I \wedge
  e^J \right) \wedge F_{J I}(\omega^{ SD}) \nonumber \\ &+&
(\alpha_3-i \alpha_4)d\sL_{CS}(\omega^{ASD})+
(\alpha_3+i \alpha_4)d\sL_{CS}(\omega^{SD}).
\ea
Notice that the momentum shift (\ref{equ})---analog of the canonical
transformation induced by the addition of the Pontrjagin invariant in
Yang-Mills theory that introduces the $\theta$ parameter in QCD---can only be
obtained for values of the free parameters in the action (\ref{action})
that make the formulation complex.

\section{The theta parameter in general relativity}\label{thetas}

In the previous section we have shown that the most general family
of connection variables that can be obtained from the standard
Ashtekar-Barbero variables and general action (\ref{action})
contains the $\theta$ parameter family only in the complex self-dual
or anti-self-dual formulations.  Therefore, contrary to what one
might have naively expected, the real connection formulation of
gravity with non trivial $\theta$ is not contained in the family of
possible phase space parametrizations stemming from (\ref{action}).
In this section we will show that the requirement of manifest {\em
Lorentz} invariance, that initially led to (\ref{action}), is too
restrictive to contain that case.

Indeed one can recover the canonical transformation if one is ready
to introduce a boundary term that violates the Lorentz gauge
symmetry. More precisely, using the additional structure provided by
the gauge condition (\ref{timegauge}), we define:
\begin{eqnarray}
  \hat{\omega}^{I J}_{\mu} = \epsilon^{I J}{}_{C D} ({\gamma}
  \omega_{\mu}^{K C} + \frac{1}{2} \epsilon^{CK}_{\ \ \ M N} \omega^{M N}_{\mu})
  n_K n^{D} ,\label{barber}
\end{eqnarray}

Note that $\hat{\omega}$ transforms as an $SO(3)$-connection for
 $SO(3)\subset SO(3,1)$ gauge transformations that leave invariant the
internal vector $n^I$. Notice that the components of (\ref{barber}) are such
that $\hat{\omega}^{i 0}_\mu=0$ for $i=1,2,3$, and
    $\hat{\omega}^{i j}_a = \epsilon^{i j k} \op_a^k$ for $a=1,2,3$
(coordinates adapted to the foliation $\Sigma$). It follows from this that the curvature components
  $\hat{F}_{\mu\nu}^{i 0}=0$, $\hat{F}_{t d}^{i j} = \partial_{[t} \hat{\omega}^{i j}_{d]} -
  \hat{\omega}^{i k}_{[t} \hat{\omega}^{k j}_{d]}$, and $\hat{F}_{a b}^{i j}=\epsilon^{i j k} F^k_{a b} [\op]$.
Now we can introduce topological boundary terms defined in terms of the
connection (\ref{barber}). Notice that the Euler term vanishes
identically. This leaves the Pontrjagin term for $\hat{\omega}^{I J}_{\mu}$
which takes the form
\begin{eqnarray}
  \alpha_7 \int _{\cal M} {\rm Tr} \left[ F(\hat{\omega})\wedge F(\hat{\omega})
  \right] = 2 \alpha_7 \epsilon^{abc} (\partial_t (\op_{a})_k F_{bc}^k(\op) +
  t^\mu (\hat \omega_\mu)_k {D}_a F^k_{bc}(\op)),
\label{DefP}
\end{eqnarray}
Notice that the second term in the previous expression vanishes due
to the Bianchi identities. The Pontrjagin term depends only on $\op$;
therefore, it satisfies the condition (\ref{clave}). In fact it is
obvious from the form of the previous expression that its effect is
the expected one producing the canonical transformation \be
\Pp^a_i=4\frac{\alpha_1}{\gamma} E^a_i+ 2 \alpha_7 \epsilon^{abc}
F_{bci}(\op), \label{equte} \ee which is real for real Immirzi
parameter $\gamma$.

We have just shown how the theta term in quantum gravity can be
obtained from the addition of the total derivative (\ref{DefP}). In
other words, the canonical transformation studied in the quantum
context in \cite{largegauge} cannot be obtained from the most
general manifestly Lorentz invariant first order formulation of
gravity. In order to define the appropriate boundary term one needs
to introduce a boundary term that brings in an $SO(3)\subset
SL(2,C)$ by explicitly choosing an internal vector $n^I$. This might
seem strange at first sight as one would seem to be violating both
Lorentz invariance in sharp conflict with general covariance.  From
the point of view of the classical theory is its clear that this is
not the case since the term added has no effect on the equation of
motion of the theory. However, the situation might appear more
obscure in the quantum theory. After all we have seen that---as in
QCD--- the theta term can have important dynamical as well as
kinematical effects in the quantum theory. So even when it is clear
that no violation of Lorentz or diffeomorphism invariance is present
in the classical theory (i.e. on shell) we need to make sure that
this remains true in the context of quantum gravity where off-shell
contributions to physical amplitudes cannot be avoided.

So, can the boundary term (\ref{DefP}) produce a Lorentz violating
effect in the quantum theory? The answer to this question is in the
negative as we argue now. The reason is the topological character of
(\ref{DefP}). The quantity computed in (\ref{DefP}) is proportional
to the Pontrjagin invariant of an $SU(2)$ principal bundle obtained
through the choice of an internal normalized vector $n^I$ (for a
mathematical description see \cite{Fatibene:2007ce}). As the latter
takes discrete values it must be invariant under continuous
deformations of $n^I$. It remains the question of whether there are
homotopically inequivalent choices of $n^I$. This correspond to the
possible winding of the maps from $\sM=\Sigma\times \R$ into the
hyperboloid $H\in \MM^4$ defined by the condition
$n^In^J\eta_{IJ}=-1$. As this winding is trivial we conclude that
the term (\ref{DefP}) is independent of the choice of $n^I$ and
hence well defined.

\section{Conclusions}\label{con}

We have completed the canonical analysis of the general action
(\ref{action}) and obtained the Dirac bracket for arbitrary values
of the couplings $\alpha_1$ to $\alpha_6$. As long as we restrict to
this action the family of connection formulations is described by
the following two cases: \begin{enumerate} \item {\em real
variables} The phase space variables are labelled by an $SU(2)$
connection is given by $\op=\Gamma+\gamma \hat K$ with Immirzi
parameter $\gamma=2\alpha_1/(2\alpha_2-\alpha_5)$ and
$\alpha_3=\alpha_4=0$, and conjugate momentum
$\Pp^a_i=4\frac{\alpha_1}{\gamma} E^a_i$. This shows that both
$\alpha_2$ and $\alpha_5$ enter the definition of the Immirzi
parameter. Hence it is possible to obtain a non trivial $\gamma$ by
simply adding the Nieh-Yang topological invariant to the Palatini
action as shown in \cite{Date:2008rb}.
\item {\em complex variables} The configuration variable is described by a self
dual or antiselfdual connection $\op=\Gamma+\gamma \hat K$ with
$\gamma=\pm i$ and the other parameters constrained to satisfy $\pm i
(2\alpha_2-\alpha_5)=2\alpha_1$, with  $\alpha_3$ and $\alpha_4$
arbitrary. The conjugate momentum is
$\Pp^a_i=4\frac{\alpha_1}{\gamma} E^a_i + 2 (\alpha_3+\gamma
\alpha_4) \epsilon^{abc} F_{bci}(\op)$. This second set contains the
one studied by \cite{merced} as a subclass.
\end{enumerate}
If in turn one wants to describe the effects of $SU(2)$ large gauge
transformations for real variables by the addition of a term to the
first order action one has no choice but to break manifest Lorentz
invariance by the addition of the term (\ref{DefP}) to the action
(\ref{action}) with parameters in the first class above. This
explicit symmetry breaking is only apparent as the term added does
not affect the classical equations of motion on the one hand, and it
does change the quantum theory but in a Lorentz invariant way as
argued in the last section.

Finally notice that if one is ready to break manifest Lorentz
invariance in a more general way then the set of possible connection formulations become
infinite dimensional. For instance the canonical transformations
\be \label{canoni}(\op_a^i,\Pp^b_j)\rightarrow (\tilde \op_a^i=\op_a^i+\delta
W[\Pp]/\delta \Pp^a_i ,\Pp^b_j)\ee for $W[\Pp]$ an arbitrary
diffeomorphism invariant and $SU(2)$ invariant functional of
$\Pp_a^i$. This kind of canonical transformation---consisting of
shifting one canonical variable by the total derivative of a
functional of the canonically conjugate one---is available in any
field theory. For instance in the case of a real scalar field $\phi$
with conjugate momentum $\pi$ then the analog of the canonical
transformations
 above is given by the shift $(\pi,\phi) \rightarrow (\pi+f(\phi),\phi)$
for some $f:\R\rightarrow\R$. The quantization in this case strongly
depends on the choice of canonical variables. Notice that this
transformation would turn a simple free theory (which can
straightforwardly quantized using for instance the Fock
representation) into a highly non linear (depending on $f(\phi)$)
theory where those techniques cannot be directly applied.

The situation in the case of LQG is much simpler at first sight.
The reason is that the canonical transformation (\ref{canoni}) preserves the
connection nature of the configuration variable and therefore allows
for a straightforward implementation of the standard LQG
quantization techniques: definition of holonomy-flux algebra of
basic kinematical observables and construction of the (unique)
diffeomorphism invariant representation.  However, this uniqueness
of the construction appears to have some unexpected implications. On
the one hand questions concerning the geometric interpretation of
the kinematical variables in the kinematical Hilbert space seem to arise,
as well as the possibility of physically distinguishable sectors
(due to the potential unitarily inequivalence of the different formulations). 
As this concerns entirely the quantum theory these questions will
be investigated elsewhere. In the appendix we explicitly exhibit the
infinite dimensional nature of this family of connection formulations.

\section{Acknowledgements}
The authors would like to thank discussions with M. Montesinos
and C. Rovelli on the subject of this work. A.P. acknowledges the support of
the Institut Universitaire de France and the grant
ANR-06-BLAN-0050 of the Agence Nationale de la Recherche.

\section{Appendix}

In order to simplify the notation let us assume that we are in the real
connection variables setting. The functional $W[\Pp]$ appearing in (\ref{canoni})
can be simply thought of a functional $W[E]$ as the triad $E$ is proportional to
$\Pp$ (at least in the $\alpha_7=0$ case).
An example of suitable generating functional is \ba && W_2[E]=\int_{\Sigma} \lambda_1
{\sL}_{CS}(\Gamma)+ \lambda_2 \sqrt {{\rm det}(E)}+\lambda_3 R[E]
\sqrt {{\rm det}(E)}+\lambda_4 R_{abcd}R^{abcd}[E] \sqrt {{\rm
det}(E)} +\cdots \label{tres}\ea where $R[E]$ and $R_{abcd}$ is the
scalar curvature and Riemann tensor of $\Sigma$ associated to $E$,
and $\sL_{CS}$ is the Chern-Simons Lagrangean evaluated in the spin
connection $\Gamma^i_a(E)$. Unlike the previous case, generating
function $W_2$ contains infinitely many parameters---we have given a
few characteristic examples; however, any scalar density local
functional of $E$ is assumed to be contained in $W_2$. The action on
the connection variables is to shift the connection as
$(A,E)\rightarrow (A+\delta W_3/\delta E,E)$. This observation
implies that an infinite dimensional set of simple connection
variables for general relativity exist.

Notice the word simple in the previous sentence. The fact that the
connection formulations infinitely dimensional should have been
expected from the fact that any phase space functional  generates a
Hamiltonian vector field which (if the latter is diffeomorphism
invariant and gauge invariant) can be viewed as a one parameter
family of canonical transformations preserving the connection nature
of the variables. However, the most general transformation is
generated by functional depending on both the connection and the
electric field and in general these transformations will not be
analytically integrable or be more complicated. Here we
concentrate on infinitesimal canonical transformations which
can be explicitly exponentiated and lead to a close formula for the
new variables as a function of the old ones.

Among the functionals of both the connection $A$ and $E$ whose
associated Hamiltonian flow can be integrated in close form there is
an important example, namely
 \ba W_3[A,E]=\epsilon \int_{\Sigma}
(A^i_a-\Gamma_a^i)E^a_i+\cdots\label{dos} , \ea The generating
function $W_2$ generates rescaling of the Immirzi parameter
$\gamma\rightarrow (1+\epsilon)\gamma$. The exponentiated version
generates finite redefinition of the Immirzi parameter. This one is
clearly not unitarily implementable at the kinematical level in LQG.

The existence of an infinite dimensional set of possible $SU(2)$
connection formulations of general relativity has little interest
from the classical point of view. They are all equivalent ways of
writing the same classical theory. However, questions arise as 
to what the interpretation and effects of these parameters might 
be in the quantum theory. These questions will be addressed elsewhere.

\end{document}